\documentclass[aps,prl,floatfix,showpacs,showkeys,twocolumn,superscriptaddress]{revtex4-1}
\usepackage{bm}
\input{epsf}

\begin{document}

\title{Bound electron pair in a MOS-structure}

\author{M.M. Mahmoodian}
 \affiliation{Rzhanov Institute of Semiconductor Physics, Siberian Branch of the Russian Academy of Sciences, \\Novosibirsk, 630090, Russia}
 \affiliation{Novosibirsk State University, Novosibirsk, 630090, Russia}

\author{A.V. Chaplik}
 \affiliation{Rzhanov Institute of Semiconductor Physics, Siberian Branch of the Russian Academy of Sciences, \\Novosibirsk, 630090, Russia}
 \affiliation{Novosibirsk State University, Novosibirsk, 630090, Russia}

\date{\today}

\begin{abstract}
In developing our previous contribution (arXiv:1804.00889) we have numerically found the bound state energy and correspondent wave function of the two electrons confined to move in a quantum well placed close to the gate electrode. Spin-orbit interaction (SOI) and image charge forces result in effective attraction between electrons. We considered also the effect of gate voltage applied to the structure and discovered that this can essentially increase the bound energy of the pair so that it remains stable even at room temperature.
\end{abstract}

\maketitle

As we have have shown in \cite{b1} for an exactly solvable but rather crude model two electrons in a 2D system with a closely placed metal electrode attract each other due to SOI and images forces. Under proper relations between the structure parameters this attraction exceeds the Coulomb repulsion and formation of the bound state-bielectron - becomes possible. Now we have considered the same problem in the exact positioning and have numerically found the bound energy of the electron pair, also with accounting for an additional electric field $\overline{F}$ created by the gate voltage. In this case the spin-orbit contribution into the Hamiltonian reads
\begin{eqnarray}\label{f1}
\hat{H}_{SO}=A\overline{F}{\bf n}\left(\left[\hat{\bf{p}}_1\bm{\sigma}_1\right]+\left[\hat{\bf{p}}_2\bm{\sigma}_2\right]\right)+\nonumber\\
+A\widetilde{F}{\bf n}\left[\left(\hat{\bf{p}}_1-\hat{\bf{p}}_2\right)\left(\bm{\sigma}_1-\bm{\sigma}_2\right)\right],
\end{eqnarray}
where ${\bf n}$ is the unit normal vector, $A$ is the constant determining the SO coupling constant $\alpha$ in the Rashba term $\alpha=A\overline{F}$. $\widetilde{F}$ is the normal to the gate component of the contribution of the interaction of a given electron with the image of another one:
\begin{eqnarray}\label{f2}
\widetilde{F}=\frac{eD}{\varepsilon\left(D^2+\rho^2\right)^{3/2}},~~~\rho=\left|\bm{\rho}_1-\bm{\rho}_2\right|.
\end{eqnarray}
Here $\varepsilon$ is the dielectric permeability, $D/2$ is the distance between quantum well and metal (see Fig.~1). 

SOI depending on $\widetilde{F}$ is written in the form $A\widetilde{F}{\bf n}\left[(\hat{\bf{p}}_1-\hat{\bf{p}}_2)(\bm{\sigma}_1-\bm{\sigma}_2)\right]$, as this energy depends on the electron velocity relative to the source of the field and that's why the fields of the own image charges of electrons do not contribute to SOI.

\begin{figure}[h]\label{fig1}
\leavevmode\centering{\epsfxsize=6.5cm\epsfbox{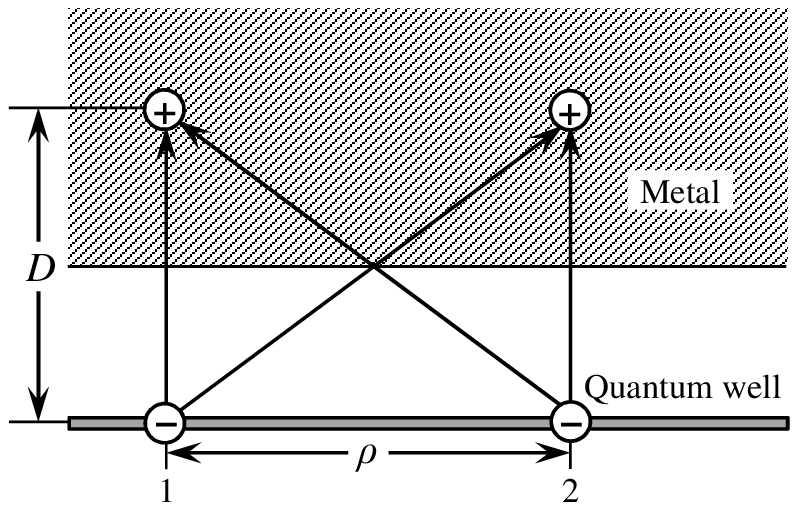}}
\caption{Sketch of the structure. The arrows show the directions of the forces acting on the electrons.}
\end{figure}

The situation becomes especially simple for a quantum wire (instead of quantum well). Take the $OX$-axis along the wire, $OY$-axis along the normal to the gate. Further we use the center-of-mass reference frame $X=(x_1+x_2)/2$, $x=x_1-x_2$ and consider a pair at rest as a whole. The system of four equations for the components of the bi-spinor wave function is now totally decoupled:
\begin{eqnarray}\label{f3}
\hat{H}_0\psi_1=E\psi_1,\nonumber\\
\hat{H}_0\psi_2-2A\left\{\hat{p}_x\mathcal{F}(x)\psi_2\right\}=E\psi_2,\nonumber\\
\hat{H}_0\psi_3+2A\left\{\hat{p}_x\mathcal{F}(x)\psi_3\right\}=E\psi_3,\\
\hat{H}_0\psi_4=E\psi_4.\nonumber
\end{eqnarray}
Here $\hat{H}_0$ contains kinetic energy and Coulomb repulsion $V_c$ of the electron pair, $\mathcal{F}=\overline{F}+2\widetilde{F}$ is given by Eq.~(\ref{f2}) with the replacement $\rho$ by $x$.

In the equations for $\psi_1$ and $\psi_4$ SOI-term is absent, $\hat{H}_0$ gives only Coulomb repulsion, so these equations do not have the localized solutions. That's why we seek for solution for the bielectron in the form $(0,\psi_2,\psi_3,0)$. We introduce the functions $\chi_2$ and $\chi_3$ in accord with
\begin{eqnarray}\label{f4}
\psi_2=\exp\left(i\int\limits_0^x\mathcal{F}(x')dx'\right)\chi_2(x),\nonumber\\
\psi_3=-\exp\left(-i\int\limits_0^x\mathcal{F}(x')dx'\right)\chi_3(x),
\end{eqnarray}
and obtain for them the identical equations:
\begin{eqnarray}\label{f5}
-\frac{1}{m}\chi_{2,3}''+\left[V_c(x)-mA^2\mathcal{F}^2(x)\right]\chi_{2,3}=E\chi_{2,3}.
\end{eqnarray}

As one sees from Eq.~{\ref{f5}} SOI, independently of the sign of the constant $A$, gives attraction between the electrons. The existence of the bound state becomes just a problem of relation between $V_c$ and $mA^2\mathcal{F}^2$. We plot the total potential $U(x)=V_c(x)-mA^2\mathcal{F}^2(x)$ and the ground state wave function in Fig.~2 for $\overline{F}=2.24\cdot10^5\mbox{V}/\mbox{cm}$, $D=30$~{\AA}, $m=0.1m_e$, $\varepsilon=10$, $A=2\cdot10^2~\mbox{cm}^2/(\mbox{V}\cdot\mbox{s})$ (for quantum wells of $Bi_2Se_3$, see \cite{b2,b3}). The dependence of the bielectron bound energy on the gate voltage both for 1D and 2D cases is depicted in Fig.~3. Remarkably that for a quite realistic value $D=50$~{\AA} and rather moderate additional field $\overline{F}=9\cdot10^5\mbox{V}/\mbox{cm}$ the bound energy of the bielectron in 2D case equals $30.1~\mbox{meV}$, that corresponds to the room temperature.

\begin{figure}[ht]\label{fig3}
\leavevmode\centering{\epsfxsize=6.5cm\epsfbox{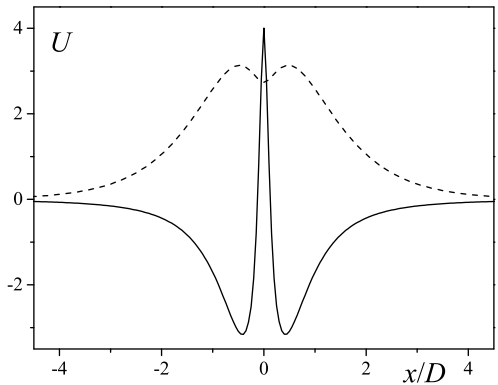}}
\caption{The dependence of the total potential $U=V_c-mA^2\mathcal{F}^2$ in quantum wire in units of $\hbar^2/(mD^2)$ (solid line) and the ground state wave function (dashed line).}
\end{figure}

\begin{figure}[ht]\label{fig4}
\leavevmode\centering{\epsfxsize=6.5cm\epsfbox{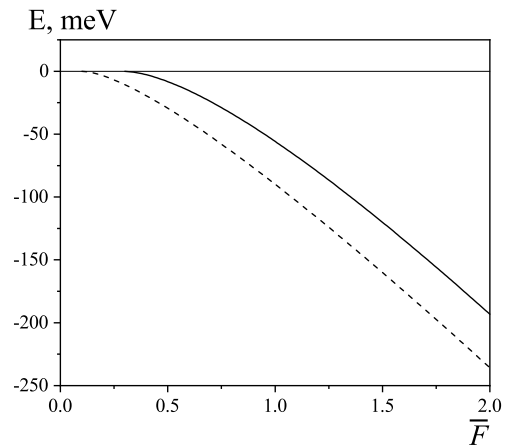}}
\caption{The dependence of the bielectron bound energy on the gate voltage in quantum wire (solid line) and 2D system (dashed line). Electric field $\overline{F}$ is measured in units of $e/\varepsilon D^2=1.6\cdot10^5~$V/cm.}
\end{figure}

One of possible experimental manifestations of the bi-electrons formation could be measurement of the quantum wire conductance $G$. As long as one deals with single electrons $G$ is given by the well known staircase-like function of the carrier concentration $n$ (e.g. quantum point contact). When pairs appear the charge carriers become bosons, the steps in $G$ should vanish and dependence $G(n)$ becomes smooth.

The extended paper in which we found also the bound energy dependence on the center-of-mass momentum is submitted to JETP.


\begin{thebibliography}{8}

\bibitem{b1} M.~M.~Mahmoodian and A.~V.~Chaplik, Pis'ma Zh. Eksp. Teor. Fiz. {\bf 107}, 590 (2018). [JETP Lett. {\bf 107}, 564 (2018)].


\bibitem{b2} A.~Manchon, H.~C.~Koo, J.~Nitta, S.~M.~Frolov and R.~A.~Duine, Nature materials {\bf 14}, 871 (2015).

\bibitem{b3} P.~D.~C.~King, R.~C.~Hatch, M.~Bianki et al, Phys. Rev. Lett. {\bf 107}, 096802 (2011).

\end{thebibliography}
\end{document}